\documentclass[11pt,a4paper]{article}
\pdfoutput=1
\usepackage{jheppub}
\usepackage{slashed}

\makeatletter
\def\@fpheader{\relax}
\makeatother

\usepackage[czech,english]{babel}
\usepackage{graphicx}
\usepackage{amsmath,amsfonts,amssymb}
\DeclareMathOperator{\MyProd}{\scalebox{1.4}{$\mathrm{I\kern-0.2ex I}$}}

\title{Note on Nonlinear Schr\"odinger Equation, KdV Equation and 2D Topological Yang-Mills-Higgs Theory}

\author[a]{Jun Nian}

\emailAdd{nian@ihes.fr}

\affiliation[a]{Institut des Hautes \'Etudes Scientifiques\\
	Le Bois-Marie, 35 route de Chartres\\
         91440 Bures-sur-Yvette, France\\}

\abstract{In this paper we discuss the relation between the (1+1)D nonlinear Schr\"odinger equation and the KdV equation. By applying the boson/vortex duality, we can map the classical nonlinear Schr\"odinger equation into the classical KdV equation in the small coupling limit, which corresponds to the UV regime of the theory. At quantum level, the two theories satisfy the Bethe Ansatz equations of the spin-$\frac{1}{2}$ XXX chain and the XXZ chain in the continuum limit respectively. Combining these relations with the dualities discussed previously in the literature, we propose a duality web in the UV regime among the nonlinear Schr\"odinger equation, the KdV equation and the 2D $\mathcal{N}=(2,2)^*$ topological Yang-Mills-Higgs theory.}

\keywords{nonlinear Schr\"odinger equation, KdV equation, Bethe Ansatz equation, topological Yang-Mills-Higgs theory, duality}

\arxivnumber{}

\newcommand{\bea}{\begin{eqnarray}}
\newcommand{\eea}{\end{eqnarray}}

\newcommand{\be}{\begin{equation}}
\newcommand{\ee}{\end{equation}}

\def\Xint#1{\mathchoice
   {\XXint\displaystyle\textstyle{#1}}%
   {\XXint\textstyle\scriptstyle{#1}}%
   {\XXint\scriptstyle\scriptscriptstyle{#1}}%
   {\XXint\scriptscriptstyle\scriptscriptstyle{#1}}%
   \!\int}
\def\XXint#1#2#3{{\setbox0=\hbox{$#1{#2#3}{\int}$}
     \vcenter{\hbox{$#2#3$}}\kern-.5\wd0}}

\def\dashint{\Xint-}

\begin{document}
\maketitle

\section{Introduction}\label{sec:introduction}

The (1+1)D nonlinear Schr\"odinger equation (NLS) and the Korteweg-de Vries equation (KdV) are two well-studied and intimately related integrable models. They share many properties both at the classical level and at the quantum level. Some previous studies have found their relation in some limits. For instance, it is known that using a certain Ansatz the nonlinear Sch\"odinger equation can be approximated by the KdV equation \cite{mathpaper}. In this paper, we would like to pursue their relation from some new perspectives, e.g. the boson/vortex duality \cite{Zee, Gubser, BEC, BECstring}.

In principle, the nonlinear Schr\"odinger equation can be generalized to the dimensions other than (1+1)D. The (3+1)D nonlinear Schr\"odinger equation is also called the Gross-Pitaevskii equation. It was shown in Refs.~\cite{Zee, Gubser} that using the boson/vortex duality one can map the Gross-Pitaevskii equation into an effective closed string theory. Recently, in Refs.~\cite{BEC, BECstring} this (3+1)D duality was generalized to include boundaries, and the dual theory is an effective open string theory. The same steps can be repeated in (1+1)D, however, instead of an effective string theory the dual theory is another scalar field theory, and the bulk equation of motion in a certain limit becomes the KdV equation. Therefore, the boson/vortex duality helps us relate two apparently different integrable models.

At quantum level, the quantum KdV equation was first studied in Ref.~\cite{Zamolod-1}, where it was shown that the quantum KdV equation can be obtained from a conformal field theory. Later, the Bethe Ansatz equation for the quantum (modified) KdV equation was found in Refs.~\cite{Kundu, Volkov, qu2KdV}, which can be viewed as the continuum limit of the spin-$\frac{1}{2}$ XXZ chain. On the other hand, it was known before that the quantum nonlinear Schr\"odinger equation can be viewed as the continuum limit of the spin-$\frac{1}{2}$ XXX chain \cite{Korepin, Faddeev}. Hence, in the small anisotropy limit of the spin-$\frac{1}{2}$ XXZ chain or equivalently in the UV regime, the Bethe Ansatz equations of the quantum KdV equation and the quantum nonlinear Schr\"odinger equation coincide.

Besides the relations among integrable models, in recently years people have also found many relations between integrable models and gauge theories. In particular, in Refs.~\cite{GS-1, GS-2} it was found that the quantum nonlinear Schr\"odinger equation is dual to the 2D $\mathcal{N}=(2,2)^*$ topological Yang-Mills-Higgs theory, which was constructed in Ref.~\cite{HiggsBundle}. Therefore, we can propose a duality web among these two theories and the KdV equation in the UV regime.

There are some alternative ways of obtaining the KdV equation. One possible way is to take the hydrodynamic limit of the elliptic Calogero model, in which the particle number $N \to \infty$. The interaction of the obtained theory has both real periods and imaginary periods. By taking different limits of the imaginary periods, one obtains the Benjamin-Ono equation (BO) and the KdV equation \cite{Abanov}. One can also consider the elliptic generalization of the BO equation, which is the intermediate long wave eqation (ILW). The Bethe Ansatz equation of the ILW equastion was studied in Ref.~\cite{Litvinov}. Recently, it was also proposed to use the Bethe/Gauge correspondence to study the ILW equation, whose dual theory is a 6-dimensional supersymmetric gauge theory \cite{Bonelli, Koroteev}. Although in this way, one can obtain the same Bethe Ansatz equation as in Ref.~\cite{Litvinov}, at the moment it is still not very clear how to obtain the quantum KdV equation by taking an appropriate limit.

This paper is organized as follows. In Section~\ref{sec:duality} we briefly review some known facts about the classical nonlinear Schr\"odinger equation and the classical KdV equation. After generalizing the boson/vortex duality to the (1+1) dimensions, we discuss that the bulk equation of motion for the dual theory becomes the classical KdV equation in a certain limit. The previously known relation between these two theories in the literature will also be recalled. In Section~\ref{sec:quKdV}, we move on to the discussion of the quantum case. We start with a review on the known facts about the quantum KdV equation. After that we discuss how the quantum KdV equation and the quantum nonlinear Schr\"odinger equation can be viewed as the continuum limits of the spin-$\frac{1}{2}$ XXZ chain and XXX chain respectively. In Section~\ref{sec:GS}, we attach the KdV equation to the previously established relation to form a duality web among the nonlineaer Schr\"odinger equation, the KdV equation and the 2D $\mathcal{N}=(2,2)^*$ topological Yang-Mills-Higgs theory. Finally, we discuss some open problems in Section~\ref{sec:discussion}.

\section{Classical NLS and KdV}\label{sec:duality}

In this section, we would like to discuss the (1+1)D classical nonlinear Schr\"odinger equation and the classical KdV equation as well as their relations. It is known previously in the mathematical literature that one can obtain the classical nonlinear Schr\"odinger equation from the classical KdV equation using a certain Ansatz. As we will discuss in this section, it turns out that the so-called boson/vortex duality can map the classical nonlinear Schr\"odinger equation back into the KdV equation in a quite unexpected way.

\subsection{Review of the Theories}

Before we discuss the relation between the classical nonlinear Schr\"odinger equation and the classical KdV equation, let us briefly review these theories in this subsection.

The (1+1)D nonlinear Schr\"odinger equation (NLS) is
\be\label{eq:NLS}
  i \partial_t \phi = -\frac{1}{2} \partial_x^2 \phi + 2 c (\phi^* \phi) \phi\, .
\ee
The Hamiltonian of the theory is given by
\be
  \mathcal{H} = \int dx \, \left[\frac{1}{2} \frac{\partial \phi^*}{\partial x} \frac{\partial \phi}{\partial x} + c \left(\phi^* \phi \right)^2 \right] \, ,
\ee
where the field $\phi$ has the Poisson structure
\be
  \{\phi^* (x),\, \phi(x') \} = \delta (x-x') \, .
\ee
In (1+1)D, this theory is integrable both at the classical level and at the quantum level.

For the attractive interaction, i.e. $c<0$, the nonlinear Schr\"odinger equation has the so-called bright soliton solution. In the literature it was known that for the attractive interaction, the quantum $N$ particles become $N$ solitons when $N$ is large \cite{Kulish, photosoliton}. Let us briefly recall some facts in the following.

A bright soliton solution to the nonlinear Schr\"odinger equation \eqref{eq:NLS} is given by
\be
  \phi = \sqrt{\frac{|c|}{2}} \, \textrm{sech} (|c| (x - x_0)) \, \textrm{exp} \left(\frac{i}{2 c^2} t \right) \, .
\ee
One can also generalize this solution to the $N$ coincident solitons, which is
\be
  \phi = N \sqrt{\frac{|c|}{2}} \, \textrm{sech} (|c| N (x - x_0)) \, \textrm{exp} \left(\frac{i N^2}{2 c^2} t \right) \, .
\ee
The general $N$-soliton solution can be found using the inverse scattering method \cite{Takhtajan}. When the $N$ solitons are well-separeted, the solution is given by \cite{Gordon}:
\be
  \phi (x, t) = \sum_{j=1}^N u_j (x, t)\, ,
\ee
where
\be
  u_j (x, t) = A_j \, \textrm{sech} \left(A_j  (x - x_j) + q_j \right) \, e^{i (\Phi_j + \psi_j)}\, ,
\ee
with $x_j = x_j^0 + p_j t$ and $\psi_j = \psi_j^0 + p_j x + t (A_j^2 - p_j^2)/2$ denoting the position and the phase of the $i$-th soliton respectively. $A_j$ is the size, while $\Phi_j$ and $q_j$ are the phase shift and the collisional position shift respectively. They satisfy
\be
  q_j + i \Phi_j = \sum_{k \neq j} \textrm{sgn} (x_k - x_j)\, \textrm{ln} \left[\frac{A_j + A_k + i (p_j - p_k)}{A_j - A_k + i (p_j - p_k)} \right]\, ,\quad \textrm{for } j = 1\, ,\, \cdots,\, N\, .
\ee

For the repulsive interaction, i.e. $c>0$, one can consider a constant background, and the nonlinear Schr\"odinger equation (NLS) \eqref{eq:NLS} is modified as
\be\label{eq:modNLS}
  i \partial_t \phi = -\frac{1}{2} \partial_x^2 \phi + 2 c (\phi^* \phi) \phi - 2 c \phi\, .
\ee
It has the so-called dark soliton solution:
\be
  \phi = \textrm{tanh} (\sqrt{2 c} x)\, .
\ee
By doing a Galilean boost, one can also obtain the moving grey soliton solution.

The Korteweg-de Vries equation (KdV) is another well-known (1+1)D integrable model. It can be written as
\be\label{eq:KdV}
  u_t + 6 u u_x + u_{xxx} = 0\, ,
\ee
where the subscripts denote the derivatives. It has the simple soliton solution:
\be
  u = \frac{k_1^2}{2} \, \textrm{sech}^2 \left[\frac{1}{2} \left(k_1 x - k_1^3 t + \eta_1^{(0)} \right) \right]\, ,
\ee
where $k_1$ and $\eta_1^{(0)}$ are constants. To find its $N$-soliton solution, one can apply the finite pole expansion \cite{KdVref, Ablowitz}:
\be
  u = -2 \sum_{i=1}^N \frac{1}{\left(x - x_i (t) \right)^2}\, .
\ee 
Plugging this Ansatz into Eq.~\eqref{eq:KdV}, one obtains
\be
  \sum_{i=1}^N \frac{-2}{(x - x_i)^3} \left[- \dot{x}_i - 12 \sum_{j=1,\, j\neq i}^N \frac{1}{(x - x_j)^2} \right] = 0\, .
\ee
By setting $x = x_I + \epsilon$ ($\epsilon \to 0$) and imposing that the terms $\sim 1 / \epsilon^3$, $\sim 1 / \epsilon^2$ vanish, one obtains
\be
  \dot{x}_I = -12 \sum_{j=1,\, j\neq I}^N \frac{1}{(x_I - x_j)^2}\, ,
\ee
and
\be
  \sum_{j=1,\, j\neq I}^N \frac{1}{(x_I - x_j)^3} = 0\, .
\ee
These equations correspond to an $N$-body system given by the Hamiltonian
\be
  H = \frac{1}{2} \sum_{i=1}^N \dot{x}_i^2 + \frac{12^2}{2} \sum_i \sum_{j\neq i} \frac{1}{(x_i - x_j)^4}\, .
\ee

Through the so-called Miura transformation, the KdV equation can also be brought into the modified KdV equation (mKdV). If we start with the KdV equation \eqref{eq:KdV}:
\be
  u_t + 6 u u_x + u_{xxx} = 0\, ,
\ee
and changing the sign $u \to -u$ leads to
\be\label{eq:KdVtemp}
  u_t + u_{xxx} = 6 u u_x\, .
\ee
The Miura transformation
\be
  u = \partial_x v + v^2
\ee
maps the special form of the KdV equation \eqref{eq:KdVtemp} into the defocusing mKdV equation:
\be
  v_t + v_{xxx} = 6 v^2 v_x\, .
\ee
A different Miura transformation
\be
  u = \partial_x v + i v^2\, ,
\ee
relates the focusing mKdV equation:
\be
  v_t + v_{xxx} = - 6 v^2 v_x
\ee
with the complex KdV equation:
\be
  u_t + u_{xxx} = 6 i u u_x\, .
\ee

\subsection{Boson/Vortex Duality}

The (3+1)D nonlinear Schr\"odinger equation is sometimes also called the Gross-Pitaevskii equation, and in principle it can be defined in dimensions other than (3+1)D. It was shown in Refs.~\cite{Zee, Gubser} that the (3+1)D Gross-Pitaevskii equation can be mapped into an effective string theory through the so-called boson/vortex duality. Recently, this duality was also generalized to include dark solitons as boundaries \cite{BEC, BECstring}.

In this section we show how to derive the dual theory action from the Gross-Pitaevskii equation in (1+1) dimensions. The steps are similar to (2+1) dimensions \cite{ZeeBook} or (3+1) dimensions \cite{Gubser}. We choose the coordinates $(t, z)$.

Let us start with the (1+1)D Gross-Pitaevskii Lagrangian:
\be\label{eq:LGP}
  \mathcal{L}_{GP} = i \phi^\dagger \partial_t \phi - \frac{1}{2m} (\partial_z \phi^\dagger) (\partial_z \phi) - \frac{g}{2} (|\phi|^2 - \rho_0)^2\, .
\ee
Varying it with respect to $\phi^\dagger$, we obtain a version of the Gross-Pitaevskii equation:
\be\label{eq:GP}
  i\, \partial_t \phi + \frac{1}{2 m} \partial_z^2 \phi - g\, (|\phi|^2 - \rho_0)\, \phi = 0\, .
\ee
Comparing this equation with Eq.~\eqref{eq:modNLS}, we can identify the coupling constants $g \sim c$. We can parametrize $\phi$ as
\be\label{eq:Param}
  \phi = \sqrt{\rho}\, e^{i \eta}\, ,
\ee
where $\eta$ is the Goldstone boson, and $\rho$ can be thought of as the Higgs boson. It is easy to derive
\begin{align}
  \phi^\dagger & = \sqrt{\rho} \, e^{-i \eta}\, ,\nonumber\\
  \partial_t \phi & = \frac{\dot{\rho}}{2\sqrt{\rho}}\, e^{i \eta} + \sqrt{\rho} \, i \, e^{i\eta} \dot{\eta}\, ,\nonumber\\
  \partial_z \phi & = \frac{1}{2 \sqrt{\rho}}\, e^{i\eta}\, (\partial_z \rho) + \sqrt{\rho}\, i\, e^{i\eta} (\partial_z \eta)\, .
\end{align}
Hence, the original Gross-Pitaevskii Lagrangian \eqref{eq:LGP} becomes
\be
  \mathcal{L} = \frac{i \dot{\rho}}{2} - \rho \dot{\eta} - \frac{\rho}{2m} (\partial_z \eta)^2 - \frac{(\partial_z \rho)^2}{8 m \rho} - \frac{g}{2} (\rho - \rho_0)^2\, .
\ee
If we drop the first term as a total derivative, and define
\begin{align}
  \mathcal{L}_1 & \equiv - \rho \dot{\eta} - \frac{\rho}{2m} (\partial_z \eta)^2\, ,\\
  \mathcal{L}_2 & \equiv - \frac{(\partial_z \rho)^2}{8 m \rho} - \frac{g}{2} (\rho - \rho_0)^2\, ,
\end{align}
then the Lagrangian can be written as
\be
  \mathcal{L} = \mathcal{L}_1 + \mathcal{L}_2\, .
\ee
Pay attention to that in Eq.~\eqref{eq:Param} the field $\eta$ takes values in $\mathbb{R} / 2 \pi \mathbb{Z}$, but now we temporarily release this condition, and it has values in $\mathbb{R}$. The constraint will be imposed later, which will bring another piece to the theory.

We see that the term $-\rho \dot{\eta}$ in $\mathcal{L}_1$ is not written in a Lorentz-invariant way. We can complete $\rho$ to a two-vector $f^\mu = (\rho, f)$, where $\rho$ is its zeroth component, and $f$ is an auxiliary field. Next, one can show the following identity:
\be
  \mathcal{L}_1 + \frac{m}{2\rho} \left(f - \frac{\rho}{m} \partial_z \eta \right)^2 = - \rho \dot{\eta} + \frac{m}{2 \rho} f^2 - f \partial_z \eta = - f^\mu \partial_\mu \eta + \frac{m}{2\rho} f^2\, ,
\ee
where
\begin{align}
  f^\mu & = (\rho, f)\, ,\\
  \partial_\mu \eta & = (\dot{\eta}, \partial_z \eta)\, .
\end{align}
If we define the path integral measure to be
\be
  \int \mathcal{D} f\, \textrm{exp} \left[i \int d^2 x\, \frac{m}{2\rho} f^2 \right] = 1\, ,
\ee
then
\begin{align}
  e^{i \int d^2 x\, \mathcal{L}_1} & = \int \mathcal{D} f\, \textrm{exp} \left[i \int d^2 x\, \left(\mathcal{L}_1 + \frac{m}{2\rho} \left(f - \frac{\rho}{m} \partial_z \eta  \right)^2 \right) \right] \nonumber\\
  {} & = \int \mathcal{D} \vec{f}\, \textrm{exp} \left[i \int d^2 x\, \left(- f^\mu \partial_\mu \eta + \frac{m}{2\rho} f^2 \right) \right]\, .\label{eq:L1part}
\end{align}
Integrating $\eta$ out, we obtain
\be
  \partial_\mu f^\mu = 0\, ,
\ee
which can be solved locally by
\be
  f^\mu = \epsilon^{\mu\nu} H_\nu
\ee
with $H = d B$. Using this expression of $f^\mu$, we obtain
\be
  \frac{m}{2\rho} f^2 = \frac{m}{2\rho} H_0^2\, .
\ee

To rewrite $\mathcal{L}_2$ we first split $B$ into the background part and the fluctuation part:
\be
  B = B^{(0)} + b\, ,
\ee
and correspondingly,
\be
  H_\nu = H_\nu^{(0)} + h_\nu\, .
\ee
Since
\be
  \rho = f^0 = \epsilon^{01} H_1 = H_1\, ,
\ee
we can rewrite $\mathcal{L}_2$ as
\be
  \mathcal{L}_2 = - \frac{(\partial_z h_1)^2}{8 m \rho} -\frac{g}{2} h_1^2\, .
\ee
Consequently,
\begin{align}
  {} & \int \mathcal{D}\rho \mathcal{D}\eta \, \textrm{exp} \left[i \int d^2 x\, (\mathcal{L}_1 + \mathcal{L}_2) \right] \nonumber\\ 
  = & \, \int \mathcal{D} B \, \textrm{exp} \left[i \int d^2 x\, \left(-\frac{g}{2} \eta^{\mu\nu} h_\mu h_\nu - \frac{(\partial_z h_1)^2}{8 m \rho} \right) \right]\, ,\label{eq:2DStringAction}
\end{align}
where
\be
  \eta^{\mu\nu} \equiv \textrm{diag} \left(\frac{m}{\rho g}, 1 \right)\, .
\ee
It was discussed in Ref.~\cite{Gubser} that, when one is interested in the IR regime of the theory, the term $\sim (\partial_z h_1)^2$ can be dropped, because it contributes to the dispersion relation only in the UV regime:
\be
  \omega^2 = c_s^2 k^2 + \frac{\sim k^4}{m^2}\, .
\ee
Hence, to study the IR physics this higher-order term can be dropped, but for our purpose we will keep this term.

Let us return to the point that the theory should be invariant under $\eta \to \eta + 2\pi$, which we have not taken into account so far. The difference comes in Eq.~\eqref{eq:L1part}. Now we cannot simply integrate out $\eta$, instead we should split $\eta$ into two pieces:
\be
  - f^\mu \partial_\mu \eta = - f^\mu \partial_\mu \eta_{\textrm{singular}} - f^\mu \partial_\mu \eta_{\textrm{smooth}}\, ,
\ee
where we can only integrate out the smooth part $\eta_{\textrm{smooth}}$, which still induces the constraint
\be
  \partial_\mu f^\mu = 0\, .
\ee
For the singular part $\eta_{\textrm{singular}}$ due to the existence of vortices, there is
\begin{align}
  - f^\mu \partial_\mu \eta_{\textrm{singular}} & = -\epsilon^{\mu\nu} \partial_\nu B\, \partial_\mu \eta_{\textrm{singular}} \nonumber\\
  {} & = B\, \epsilon^{\mu\nu} \partial_\mu \partial_\nu \eta_{\textrm{singular}} \nonumber\\
  {} & = 2 \pi B \, \delta^2 \left(X^\mu - X^\mu (\tau, \sigma) \right) \, ,
\end{align}
with $X^\mu$ denoting the position of the singularity or the vortex. Formally, one can write the integral of the equation above as
\be
  - \int d^2 x\, f^\mu \partial_\mu \eta_{\textrm{singular}} = \mu_1 \int d^2 x\, B \, \delta^2 (X^\mu - X^\mu (\tau, \sigma)) = \mu_1 \int_{\Sigma_\alpha} B\, .
\ee
with $\mu_1 = 2\pi$. Therefore, in the IR regime the theory can be approximately written as
\be\label{eq:appSeff}
  \int \mathcal{D} B\, \textrm{exp} \left[i \int d^2 x\, \left(-\frac{g}{2} \eta^{\mu\nu} h_\mu h_\nu\right) + i \mu_1 \int_{\Sigma_\alpha} B \right]\, .
\ee
In order to perform calculations that can be compared with real systems, one can also add a string tension term like in Refs.~\cite{Zee, Gubser}.

The Gross-Pitaevskii equation, or the nonlinear Sch\"odinger equation, is an integrable model in (1+1)D. Hence, we expect the integrability also on the dual model side. Therefore, we would like to study the full dual theory without neglecting the term $\sim (\partial_z h_1)^2$ as in Eq.~\eqref{eq:appSeff}, and the full effective action is
\be\label{eq:fullSeff}
  S_{\textrm{eff}} = \int d^2 x\, \left[\mu_1 B(x) \, \delta^2 (x^\mu - x_0^\mu) - \frac{g}{2} \left(- \frac{m}{\rho g} (\partial_0 B)^2 + (\partial_1 b)^2 \right) - \frac{(\partial_1^2 b)^2}{8 m \rho} \right]\, .
\ee
We will derive the bulk equation of motion and see its classical integrability in the next subsection.

\subsection{Classical Integrability of the Dual Model}

Since the (1+1)D nonlinear Schr\"odinger equation is an integrable model both at classical level and at quantum level, we expect the dual model obtained from the boson/vortex duality map should also be integrable at least at classical level. In this subsection, we discuss the classical integrability of the dual model. As we will see, in the weak coupling limit the dual model is in fact the classical KdV theory written in a less familiar form.

As we discussed in the previous subsection, after the boson/vortex duality map we obtain a new effective action \eqref{eq:fullSeff}:
\begin{displaymath}
  S_{\textrm{eff}} = \int d^2 x\, \left[\mu_1 B(x) \, \delta^2 (x^\mu - x_0^\mu) - \frac{g}{2} \left(- \frac{m}{\rho g} (\partial_0 B)^2 + (\partial_1 b)^2 \right) - \frac{(\partial_1^2 b)^2}{8 m \rho} \right]\, .
\end{displaymath}
We can collect all the terms depending on the fluctuation $b$:
\begin{align}
  \mathcal{L}_{\textrm{eff}} & \supset \frac{1}{\rho_0 + \partial_1 b} \left [\frac{m}{2} (\partial_0 b)^2 - \frac{g}{2} (\rho_0 + \partial_1 b) (\partial_1 b)^2 - \frac{1}{8 m} (\partial_1^2 b)^2 \right] \nonumber\\
  {} & = \frac{1}{\rho_0\, \ell + \ell\, \partial_1 b} \left [\frac{m \ell}{2} (\partial_0 b)^2 - \frac{g}{2} (\rho_0\, \ell + \ell\, \partial_1 b) (\partial_1 b)^2 - \frac{\ell}{8 m} (\partial_1^2 b)^2 \right]\, ,
\end{align}
where $\ell$ is a constant with the dimension of length. We choose $\ell$, such that
\be
  g\, \rho_0\, \ell = 1\quad \Rightarrow\quad \rho_0\, \ell = \frac{1}{g}\, .
\ee
Consequently, the terms depending on $b$ become
\be
  \mathcal{L}_{\textrm{eff}} \supset \frac{1}{g^{-1} + \ell\, \partial_1 b} \left[\frac{m \ell}{2} (\partial_0 b)^2 - \frac{1}{2} \left(1 + \frac{1}{\rho_0} \partial_1 b \right) (\partial_1 b)^2 - \frac{\ell}{8 m} (\partial_1^2 b)^2 \right] \, .\label{eq:temp-1}
\ee
In the weak coupling limit $g \to 0$, the coefficient outside the bracket in the expression above is approximately $g$, and the terms in Eq.~\eqref{eq:temp-1} become
\be
  \mathcal{L}_{\textrm{eff}} \supset g \left[\frac{m \ell}{2} (\partial_0 b)^2 - \frac{1}{2} \left(1 + \frac{1}{\rho_0} \partial_1 b \right) (\partial_1 b)^2 - \frac{\ell}{8 m} (\partial_1^2 b)^2 \right] \, .
\ee
From these terms we can derive the equation of motion for $b$ in the weak coupling limit ($g \to 0$):
\be\label{eq:temp-2}
  m \ell\, \partial_0^2 b = \partial_1^2 b + \frac{3}{\rho_0} (\partial_1 b) (\partial_1^2 b) - \frac{\ell}{4 m} \partial_1^4 b\, .
\ee
We then absorb the coefficient $m \ell$ into the definition of the time $t$, morover, we can introduce some dimensionful constants on both sides to make the variables $(t, x)$ as well as the coefficients in front of eacth term dimensionless. For simplicity, we do not write these constants explicitly, but from now on all the parameters become dimensionless.

Eq.~\eqref{eq:temp-2} can be rewritten as
\be\label{eq:temp-3}
  \partial_0^2 b = \partial_1^2 b + 2 \kappa \xi (\partial_1 b) (\partial_1^2 b) + \frac{\xi}{12} \partial_1^4 b\, ,
\ee
where
\be
  \xi \equiv -\frac{3 \ell}{m}\, ,\quad \kappa \equiv - \frac{m}{2 \rho_0 \ell}\, .
\ee
As discussed above, we can introduce dimensionful constants such that the dimensionless parameter $|\xi| \ll 1$. Eq.~\eqref{eq:temp-3} is a Boussinesq-type equation. If we define the following new variables:
\be
  X \equiv x + t\, ,\quad T \equiv \kappa \xi t\, ,
\ee
then Eq.~\eqref{eq:temp-3} can be further brought into the form
\be
  b_{XT} = b_X b_{XX} + \frac{1}{24 \kappa} b_{XXXX} + o (\xi)\, ,
\ee
where the subscripts denote the derivatives. At the leading order in $\xi$, the equation above is just the KdV equation for $b_X$. By rescaling the variables $u \equiv b_X$, $X$ and $T$, we can bring this equation into the standard form of the KdV equation given by Eq.~\eqref{eq:KdV}. Hence, we obtain the (1+1)D KdV equation after applying the boson/vortex duality map to the (1+1)D nonlinear Schr\"odinger equation in the weak coupling limit ($g \to 0$), and consequently the dual theory is classically integrable in this limit.

The result that in the weak coupling limit ($g \to 0$) the nonlinear Schr\"odinger equation can be mapped into the KdV equation through the boson/vortex duality is also consistent with the previous results in the literature. For instance, for the quantum nonlinear Schr\"odinger equation given by the Hamiltonian (also called the Lieb-Lininger model):
\be
  H = -\frac{\hbar^2}{2m} \sum_{i=1}^N \frac{\partial^2}{\partial x_i^2} + g \sum_{i < j} \delta (x_i - x_j)
\ee
one can define a dimensionless coupling $\gamma$
\be
  \gamma = \frac{m}{\hbar^2 \rho_0} g\, .
\ee
Ref.~\cite{Abanov-3} has demonstrated that in the weak coupling ($\gamma \ll 1$) and weak dissipation limit the system is described by the KdV equation.

We should emphasize that in this paper we only consider the dual model in the weak coupling limit ($g \to 0$). However, the integrability should hold in the full dual model without approximation.\footnote{The author would like to thank Alexander Abanov for commenting on this point.} Moreover, we only discussed the bulk equation of motion of the dual model, while in the presence of boundaries the dual model should be described by a matrix model. These perspectives of the dual model need to be investigated in the future research.

\subsection{KdV as an Approximation of NLS}

It was studied in mathematical literature \cite{mathpaper} that the KdV equation can also be viewed as an approximation of the nonlinear Schr\"odinger equation. We briefly review the results of Ref.~\cite{mathpaper} in this subsection.

Let us start with the KdV equation \eqref{eq:KdV}
\begin{displaymath}
  \partial_t u + 6 u \partial_x u + \partial_x^3 u = 0\, .
\end{displaymath}
Next, one can adopt the following Ansatz:
\begin{align}
  u_{NLS} & = \epsilon A_1^0 (X, T) \, e^{i (k_0 x + \omega_0 t)} + \epsilon A_{-1}^0 (X, T) \, e^{-i (k_0 x + \omega_0 t)} \nonumber\\
  {} & \quad + \epsilon^2 A_2^0 (X, T) \, e^{2 i (k_0 x + \omega_0 t)} + \epsilon^2 A_{-2}^0 (X, T) \, e^{- 2 i (k_0 x + \omega_0 t)} + \epsilon^2 A_0^0 (X, T)\, ,
\end{align}
where
\be
  T = \epsilon^2 t\, ,\quad X = \epsilon (x + 3 k_0^2 t)\, ,
\ee
\be
  \omega_0 = k_0^3\, ,\quad k_0 \neq 0\, ,\quad 0 < \epsilon \ll 1\, ,
\ee
and
\be
  A (X, T) \in \mathbb{C}\, ,\quad A_{-j}^0 = \bar{A}_j^0\, .
\ee
Plugging this Ansatz into Eq.~\eqref{eq:KdV}, one obtains the following equations:
\begin{align}
  k_0^2 \, A_0^0 & = -2 A_1^0\, A_{-1}^0\, ,\\
  \partial_T A_1^0 & = - 3 i k_0 \partial_X^2 A_1^0 - 6 i k_0 (A_2^0\, A_{-1}^0 + A_1^0\, A_0^0)\, ,\\
  k_0^2\, A_2^0 & = (A_1^0)^2\, .
\end{align}
After getting rid of $A_0^0$ and $A_2^0$ from the equations above, one finally obtains the equation for $A_1^0$:
\be\label{eq:mappedNLS}
  i\, \partial_T A_1^0 = 3 k_0 \partial_X^2 A_1^0 - \frac{6}{k_0} A_1^0 |A_1^0|^2\, ,
\ee
which is the (1+1)D nonlinear Schr\"odinger equation. After redefining the variables and the coefficients, one can bring the equation above into the standard form \eqref{eq:NLS}.

From the discussions in the previous subsection, we have seen that the nonlinear Schr\"odinger equation can be mapped into a KdV equation in the weak coupling limit ($g \to 0$). The corresponding coupling in Eq.~\eqref{eq:mappedNLS} is $g \sim k_0^{-1}$, which means that the weak coupling limit ($g \to 0$) of the nonlinear Schr\"odinger equation corresponds to the UV regime, i.e. large $k_0$ or $\omega_0$, in order for the Ansatz to be valid. Hence, we expect the equivalence of the nonlinear Schr\"odinger equation and the KdV equation in the UV regime.

In Refs.~\cite{GS-1, GS-2} it was shown that the quantum nonlinear Schr\"odinger equation is dual to the 2D topological Yang-Mills-Higgs theory. Hence, combining with the previous discussions, we expect a duality web in the UV regime among the quantum nonlinear Schr\"odinger equation, the quantum KdV equation and the 2D topological Yang-Mills-Higgs theory. We will discuss the quantum KdV equation in Section~\ref{sec:quKdV} and some aspects of the duality web in Section~\ref{sec:GS}.

\section{KdV Equation}\label{sec:quKdV}

In this section, we first present some aspects of the quantum KdV equation, which was studied in Ref.~\cite{Zamolod-1} and later in some other works \cite{Volkov, Kundu, qu2KdV}. Second, we discuss the relations of the KdV equation (both classical and quantum) with some other theories, e.g. Calogero-Sutherland model and ILW equation.

\subsection{Quantum KdV Equation and Bethe Ansatz Equation}

As discussed in Refs.~\cite{Zamolod-1, qu2KdV}, it is more convenient to quantize the classical KdV equation from its relation with conformal field theory.

For a general conformal field theory, the energy momentum tensor can be expanded as
\be
  T(y) = - \frac{c}{24} + \sum_{n = -\infty}^\infty L_{-n}\, e^{i n y}\, ,
\ee
where $c$ is the central charge, and the operators $L_n$ satisfy the Virasoro algebra:
\be\label{eq:Virasoro}
  [L_m,\, L_n] = (m - n) L_{m + n} + \frac{c}{12} (m^3 - m) \, \delta_{m+n, 0}\, .
\ee
In the classical limit ($c \to -\infty$), one can make the following replacements:
\be
  T(y) \to -\frac{c}{6} u(y)\, ,\quad [\phantom{!}, \phantom{!}] \to \frac{6 \pi}{i c} \{\phantom{!}, \phantom{!}\}\, .
\ee
Consequently, the Virasoro algebra becomes the second Hamiltonian structure of the KdV equation:
\be
  \{u(y),\, u(z) \} = 2 \left(u(y) + u(z) \right)\, \delta' (y - z) + \delta''' (y - z)\, .
\ee

For the quantum KdV equation one can apply the Feigin-Fuchs transformation, which is the quantum counter-part of the Miura transformation:
\be\label{eq:FeiginFuchs}
  - \beta^2 T(y) = \, : \varphi' (y)^2 : + (1 - \beta^2)\, \varphi'' (y) + \frac{\beta^2}{24}\, ,
\ee
where
\be
  \beta = \sqrt{\frac{1-c}{24}} - \sqrt{\frac{25-c}{24}}\, .
\ee
$\varphi(y)$ can be expressed in terms of free field operators:
\be
  \varphi(y) = i Q + i P y + \sum_{n \neq 0} \frac{a_{-n}}{n} \, e^{i n y}\, ,
\ee
and the operators satisfy the Heisenberg algebra:
\be
  [Q,\, P] = \frac{i}{2} \beta^2\, ,\quad [a_n,\, a_m] = \frac{n}{2} \beta^2\, \delta_{n+m, 0}\, ,
\ee
\be
  [Q,\, a_n] = 0\, ,\quad [P,\, a_n] = 0\, .
\ee

One can define the following operator for the quantum KdV equation:
\be
  {\bf L}_j (\lambda) = \pi_j \left[e^{i \pi P H}\, \mathcal{P}\, \textrm{exp} \left(\lambda \int_0^{2 \pi} dy\, \left(: e^{-2 \varphi (y)}: \, q^{\frac{H}{2}} E + : e^{2 \varphi(y)}:\, q^{-\frac{H}{2}} F \right) \right) \right]\, ,
\ee
where $E$, $F$ and $H$ now are the generating elements of the quantum universal enveloping algebra $U_q (sl(2))$:
\be
  [H,\, E] = 2 E\, ,\quad [H,\, F] = -2 F\, ,\quad [E,\, F] = \frac{q^H - q^{-H}}{q - q^{-1}}
\ee
with $q = e^{i \pi \beta^2}$. The quantum ${\bf R}$-matrix is then determined by the quantum Yang-Baxter equation
\be
  {\bf R}_{j j'} (\lambda \mu^{-1}) \left({\bf L}_j (\lambda) \otimes 1 \right) \left(1 \otimes {\bf L}_{j'} (\mu) \right) = \left(1 \otimes {\bf L}_{j'} (\mu) \right) \left({\bf L}_j (\lambda) \otimes 1 \right) {\bf R}_{j j'} (\lambda \mu^{-1})\, .
\ee
The transfer matrix can be defined as
\be
  {\bf T}_j (\lambda) = \textrm{Tr}_{\pi_j} \left(e^{i \pi P H}\, {\bf L}_j (\lambda) \right)\, ,
\ee
which satisfies
\be
  [{\bf T}_j (\lambda),\, {\bf T}_{j'} (\mu)] = 0\, ,
\ee
and also the following functional relation
\be
  {\bf T}_j (q^{\frac{1}{2}} \lambda) \, {\bf T}_j (q^{-\frac{1}{2}} \lambda) = 1 + {\bf T}_{j - \frac{1}{2}} (\lambda) \, {\bf T}_{j + \frac{1}{2}} (\lambda)\, ,
\ee
which is the same for the integrable XXZ model.

Using the Feigin-Fuchs transformation one can map the quantum KdV equation into the quantum mKdV equation, whose algebraic Bethe Ansatz solution was presented in Ref.~\cite{Kundu}.\footnote{The Bethe Ansatz equation of a toy model of quantum KdV equation was studied in Ref.~\cite{Volkov}, and the Bethe Ansatz equations for a left and right coupled quantum KdV system were discussed in Ref.~\cite{qu2KdV}.} Following the discussions in Ref.~\cite{Kundu}, the Bethe Ansatz equation for the lattice regularized mKdV equation is given by
\be
  q^{- \kappa + L - 4 N} \left(\frac{\textrm{sin} (\lambda_j + \frac{i \eta}{2})}{\textrm{sin} (\lambda_j - \frac{i \eta}{2})} \right)^L = \prod_{\overset{k = 1}{k \neq j}}^N \frac{\textrm{sin} (\lambda_j - \lambda_k + i \eta)}{\textrm{sin} (\lambda_j - \lambda_k - i \eta)}\, ,\label{eq:BAEmKdV}
\ee
where $q \equiv e^{- \eta}$, $L$ and $N$ are the site number and the particle number respectively, and $\kappa$ appears in the generating function of the conserved quantities
\be
  \tau (\lambda) = \textrm{tr} \left(q^{-\frac{\kappa}{2} \sigma_3} {\bf T}_{\frac{1}{2}} (\lambda) \right)\, .
\ee
Eq.~\eqref{eq:BAEmKdV} is the same as the Bethe Ansatz equation for the spin-$\frac{1}{2}$ XXZ chain model with twisted boundary condition up to a factor \cite{Kundu}:
\be
  q^{- \kappa} \left(\frac{\textrm{sin} (\lambda_j + \frac{i \eta}{2})}{\textrm{sin} (\lambda_j - \frac{i \eta}{2})} \right)^L = \prod_{\overset{k = 1}{k \neq j}}^N \frac{\textrm{sin} (\lambda_j - \lambda_k + i \eta)}{\textrm{sin} (\lambda_j - \lambda_k - i \eta)}\, .\label{eq:BAEXXZ}
\ee
The spin-$\frac{1}{2}$ XXZ chain mode is defined by the Hamiltonian
\be
  H_{XXZ} = J \sum_{n=1}^N \left(\sigma_n^1 \, \sigma_{n+1}^1 + \sigma_n^2 \, \sigma_{n+1}^2 + \textrm{cos} (\eta)\, \sigma_n^3 \, \sigma_{n+1}^3 \right)\, .
\ee
We see that in the limit $\eta \to 0$ or equivalently $q \to 1$, the Bethe Ansatz equations \eqref{eq:BAEmKdV} and \eqref{eq:BAEXXZ} coincide, and the spin-$\frac{1}{2}$ XXZ chain model becomes the spin-$\frac{1}{2}$ XXX chain model:
\be
  H_{XXX} = \sum_{n=1}^N \left(\sigma_n^1 \, \sigma_{n+1}^1 + \sigma_n^2 \, \sigma_{n+1}^2 + \sigma_n^3 \, \sigma_{n+1}^3 \right)\, .
\ee
Moreover, it is known that the lattice nonlinear Schr\"odinger equation can be viewed as a generalized spin-$\frac{1}{2}$ XXX chain model \cite{Korepin, Faddeev}. Hence, we expect that in the small anisotropy limit $\eta \to 0$, the Bethe Ansatz equation of the continuous mKdV equation coincides with the one of the continuous nonlinear Schr\"odinger equation:
\be
  e^{2 \pi i \lambda_j} \prod_{k\neq j} \frac{\lambda_k - \lambda_j + i c}{\lambda_k - \lambda_j - i c} = 1\, ,\quad k = 1,\, \cdots,\, N.
\ee
In this case, the parameter $\eta$ becomes the coupling constant $c$ in the (1+1)D nonlinear Schr\"odinger equation \eqref{eq:NLS} or \eqref{eq:modNLS}, or equivalently the coupling constant $g$ in the (1+1)D Gross-Pitaevskii equation \eqref{eq:GP}. Hence, the small anisotropy limit $\eta \to 0$ corresponds to the weak coupling limit $(g \sim c) \to 0$ for the nonlinear Schr\"odinger equation, which consequently corresponds to the UV regime based the discussions in Section~\ref{sec:duality}.

\subsection{KdV Equation, Calogero-Sutherland Model and ILW Equation}

There is an alternative way of obtaining the KdV equation from the Calogero-Sutherland model and the related the intermediate long wave equation (ILW), which was discussed in Ref.~\cite{Abanov}. We briefly review this perspective of the KdV equation in this subsection.

As discussed in Ref.~\cite{Abanov}, for the Calogero-Sutherland model (CSM) given by the Hamiltonian:
\be
  H_{CSM} = \frac{1}{2} \sum_{j=1}^N p_j^2 + \frac{g^2}{2} \left(\frac{\pi}{L} \right)^2 \sum_{\overset{j, k =1}{j \neq k}}^N \frac{1}{\textrm{sin}^2 \frac{\pi}{L} (x_j - x_k)}\, .
\ee
One can define the complex coordinates $\omega_j (t) = e^{i \frac{2 \pi}{L} x_j(t)}$ and the auxiliary coordinates $u_j (t) = e^{i \frac{2 \pi}{L} y_j(t)}$, where $x_j(t)$ and $y_j(t)$ denote real coordinates and auxiliary complex coordinates respectively. These coordinates determine two new functions:
\begin{align}
  u_1 (\omega) & = g \frac{\pi}{L} \sum_{j=1}^N \frac{\omega + \omega_j}{\omega - \omega_j} = -i g \sum_{j=1}^N \frac{\pi}{L} \textrm{cot} \frac{\pi}{L} (x - x_j)\, ,\\
  u_0 (\omega) & = - g \frac{\pi}{L} \sum_{j=1}^N \frac{\omega + u_j}{\omega - u_j} = i g \sum_{j=1}^N \frac{\pi}{L} \textrm{cot} \frac{\pi}{L} (x - y_j)\, ,
\end{align}
where $\omega = e^{i \frac{2 \pi}{L} x}$. It turns out that these new functions satisfy the bidirectional Benjamin-Ono equation (2BO):
\be\label{eq:2BO}
  u_t + \partial_x \left(\frac{1}{2} u^2 + i \frac{g}{2} \partial_x \widetilde{u} \right) = 0\, ,
\ee
where
\be
  u = u_0 + u_1\, ,\quad \widetilde{u} = u_0 - u_1\, .
\ee
The functions $u_0$ and $u_1$ obey the analyticity conditions
\begin{align}
  u_1 (x) & \textrm{ analytic for } \textrm{Im} (x) \neq 0\, ;\nonumber\\
  u_0 (x) & \textrm{ analytic for } |\textrm{Im} (x)| < \epsilon \textrm{ with } \epsilon > 0\, ,
\end{align}
and the reality condition
\be
  \overline{u_1 (x)} = - u_1 (\bar{x})\, .
\ee

Assume that $\rho (x)$ and $\theta (x)$ are the density field and the velocity field respectively in the thermodynamic limit $N \to \infty,\, L \to \infty,\, N / L = const$. In terms of $\rho$ and $v = g \partial_x \theta$, one can define the right-handed and the left-handed chiral fields:
\be
  J_{R, L} = v \pm g \left[\pi \rho + \partial_x (\textrm{log} \sqrt{\rho})^H \right]\, ,
\ee
where the superscript $H$ denotes the Hilbert transform
\be
  f^H (x) = \dashint_0^L \frac{dy}{L} f(y) \, \textrm{cot} \frac{\pi}{L} (y-x)\, .
\ee
By setting one of the chiral fields to be constant, e.g. $J_L = -\pi g \rho_0$, one obtains from the 2BO equation the nonlinear chiral equation (NLC):
\be\label{eq:NLC}
  \rho_t + g \left[\rho \left(\pi (\rho - \rho_0) + \partial_x (\textrm{log} \sqrt{\rho})^H \right) \right]_x = 0\, .
\ee
On the other hand, with the boundary conditions
\begin{align}
  u(x - i 0) & = -J_0 + 2 g \left[\pi \rho + i \partial_x (\textrm{log} \sqrt{\rho})^+ \right]\, ,\nonumber\\
  \widetilde{u} (x - i 0) & = -J_0 - i u^H (x - i 0)\, ,
\end{align}
the 2BO equation \eqref{eq:2BO} becomes
\be
  u_t + \partial_x \left[\frac{1}{2} u^2 + \frac{g}{2} \partial_x u^H \right] = 0\, ,
\ee
which is equivalent to the NLC equation, and becomes the conventional Benjamin-Ono equation (BO) when the deviation of the density is smaller than the average density, i.e. $|\rho - \rho_0| \ll \rho_0$. Therefore, the NLC equation is a finite amplitude extension of the BO equation, while the 2BO equation is an integrable bidirectional finite amplitude extension of the BO equation.

The discussions above can be generalized to the elliptic Calogero model, where the interaction between particles is the Weierstrass $\wp (x | \omega_1, \omega_2)$-function with a purely real period $\omega_1$ and a purely imaginary period $i \omega_2$. The Hilbert transform is taken with respect to a strip $0 < \textrm{Im} x < \omega$ with an imaginary period $\omega$:
\be\label{eq:HilbertTrafo}
  f^H (x) = \dashint dx'\, \frac{1}{\omega_2} \, \textrm{coth} \frac{1}{\omega_2} (x - x') f(x')\, .
\ee
In the limit $\omega_2 \to \infty$, the Weierstrass $\wp$-function becomes $1 / \textrm{sin}^2 (x / \omega_1)$, and consequently the elliptic model becomes the trigonometric model. In the limit $\omega_2 \to 0$, the interaction becomes $\omega_2 \wp (x) \to \delta (x)$, and the Hilbert transform \eqref{eq:HilbertTrafo} becomes $f^H \to \omega \partial_x f$, which is local. It turns out that, in the limit $\omega_2 \to 0$, the BO equation flows to the KdV equation, while the 2BO equation flows to the nonlinear Sch\"odinger equation, which provides another link between these two equations. Moreover, the elliptic generalizations of the 2BO and the BO equation are the bidirectional intermediate long wave equation (2ILW) and the intermediate long wave equation (ILW) respectively. In particular, the ILW equation is given by \cite{ILWmathBook}:
\be
  u_t = 2 u u_x + \frac{1}{\delta} u_x + \frac{1}{\pi} \left(\partial_x^2 u\right)^H\, .
\ee
where the Hilbert transform is given by Eq.~\eqref{eq:HilbertTrafo}, and the parameter $\delta$ is proportional to the imaginary period $\omega_2$ in the elliptic Calogero model in the following way:
\be
  \delta = \frac{\pi}{2}\, \omega_2\, .
\ee
When $\delta \to \infty$, the ILW equation becomes the BO equation, while in the limit $\delta \to 0$ the ILW equation becomes the KdV equation.

For the corresponding quantum version of the theories, the Bethe Ansatz equation of the ILW equation was studied in Ref.~\cite{Litvinov} and also through the Bethe/Gauge correspondence in Ref.~\cite{Bonelli}. Recently, it was generalized to the finite difference case $\Delta ILW$ \cite{Koroteev}, which can be viewed as the hydrodynamic limit of the elliptic Ruijsenaars-Schneider model. However, as discussed in Refs.~\cite{Litvinov, Koroteev}, for the moment it is still not very clear how to obtain the Bethe Ansatz equation of the quantum KdV equation from the one of the ILW equation by taking an appropriate limit. Although there are some proposals, it is still an open problem under research.

\section{Duality Web}\label{sec:GS}

In this section, we first review the duality between the 2D $\mathcal{N}=(2,2)^*$ $U(N)$ topological Yang-Mills-Higgs theory and the (1+1)D nonlinear quantum nonlinear Schr\"odinger equation \cite{GS-1, GS-2}, and then we propose a new duality web among these two theories and the quantum KdV equation in the UV regime.

\subsection{Review of the Gerasimov/Shatashvili Duality}

The 2D $\mathcal{N}=(2,2)^*$ $U(N)$ topological Yang-Mills-Higgs theory (TYMH) was first constructed in Ref.~\cite{HiggsBundle}. It is defined by the path integral
\be
  Z_{YMH} (\Sigma_h) = \frac{1}{\textrm{Vol} (\mathcal{G}_{\Sigma_h})} \int D \varphi_0\, D \varphi_\pm\, DA\, D\Phi\, D\psi_A\, D\psi_\Phi\, D\chi_\pm \, e^S\, ,
\ee
where
\be
  S = S_0 + S_1
\ee
with
\begin{align}
  S_0 & = \frac{1}{2\pi} \int_{\Sigma_h} d^2 z\, \bigg[\textrm{Tr} \left(i \varphi_0 (F(A) - \Phi \wedge \Phi) - c \, \Phi \wedge * \Phi\right) + \varphi_+ \nabla_A^{(1,0)} \Phi^{(0,1)} \nonumber\\
  {} & \qquad\qquad\qquad\quad + \varphi_- \nabla_A^{(0,1)} \Phi^{(1,0)} \bigg] \, ,\\
  S_1 & = \frac{1}{2\pi} \int_{\Sigma_h} d^2 z\, \textrm{Tr} \bigg[\frac{1}{2} \psi_A \wedge \psi_A + \frac{1}{2} \psi_\Phi \wedge \psi_\Phi + \chi_+ \left[\psi_A^{1,0},\, \Phi^{(0,1)} \right] \nonumber\\
  {} & \qquad \qquad \quad + \chi_- \left[\psi_A^{(0,1)},\, \Phi^{(1,0)} \right] + \chi_+ \nabla_A^{(1,0)} \psi_\Phi^{(0,1)} + \chi_- \nabla_A^{(0,1)} \psi_\Phi^{(1,0)} \bigg)\, .
\end{align}
In the absence of the deformation term $c\, \textrm{Tr} (\Phi \wedge * \Phi)$ the theory preserves $\mathcal{N}=(4,4)$ supersymmetry. For generic values of $c \neq 0$ the theory preserves $\mathcal{N}=(2,2)$ supersymmetry, and the $\mathcal{N}=(2,2)$ supersymmetry transformations are given by
\be
  Q A = i \psi_A\, ,\quad Q \psi_A = - D \varphi_0\, ,\quad Q \varphi_0 = 0\, ,\label{eq:SUSY-1}
\ee
\be
  Q \Phi = i \psi_\Phi\, ,
\ee
\be
  Q \psi_\Phi^{(1,0)} = [\Phi^{(1,0)},\, \varphi_0] + c \Phi^{(1,0)}\, ,\quad Q \psi_\Phi^{(0,1)} = [\Phi^{(0,1)},\, \varphi_0] + c \Phi^{(0,1)}\, ,
\ee
\be
  Q \chi_\pm = i \varphi_\pm\, ,\quad Q \varphi_\pm = [\chi_\pm,\, \varphi_0] \pm c \chi_\pm \, .\label{eq:SUSY-4}
\ee
This theory can also be understood as the dimensional reduction of the 4D topologically twisted $\mathcal{N}=2$ $U(N)$ super Yang-Mills theory with a deformation term.

Using the technique of cohomological localization, one can compute the partition function of the 2D $\mathcal{N}=(2,2)^*$ $U(N)$ topological Yang-Mills-Higgs theory exactly:
\be
  Z_{YMH} (\Sigma_h) = e^{(1-h)\, a(c)} \sum_{\lambda \in \mathcal{R}_N} D_\lambda^{2-2h} \, e^{-\sum_{k=1}^\infty t_k\, p_k (\lambda)} \, ,
\ee
where $p_k (\lambda)$ is defined as
\be
  \frac{1}{(2 \pi)^k} \, \textrm{Tr}\, \varphi^k \, \Psi_\lambda (x_1,\, \cdots,\, x_N) = p_k (\lambda) \, \Psi_\lambda (x_1,\, \cdots,\, x_N)\, ,
\ee
and the factor $D_\lambda$ is given by
\be
  D_\lambda = \mu(\lambda)^{-1/2} \prod_{i<j} (\lambda_i - \lambda_j) \, \left(c^2 + (\lambda_i - \lambda_j)^2 \right)^{1/2} \, ,
\ee
while $\mathcal{R}_N$ denotes the set of $\lambda_i$'s satisfying the Bethe Ansatz equation:
\be
  e^{2 \pi i \lambda_j} \prod_{k\neq j} \frac{\lambda_k - \lambda_j + i c}{\lambda_k - \lambda_j - i c} = 1\, ,\quad k = 1,\, \cdots,\, N.
\ee

For the (1+1)D quantum nonlinear Schr\"odinger equation, if we consider the $N$-particle sector in the domain $x_1 \leq x_2 \leq \cdots \leq x_N$, the $N$-particle wave function satisfies the equation
\be
  \left(-\frac{1}{2} \sum_{i=1}^N \frac{\partial^2}{\partial x_i^2} \right) \Phi_\lambda (x) = 2 \pi^2 \left(\sum_{i=1}^N \lambda_i^2 \right) \Phi_\lambda (x)\, ,
\ee
where $\lambda_i$ denotes the momentum of the $i$-th particle, which satisfies the same Bethe Ansatz equation:
\be
  e^{2 \pi i \lambda_j} \prod_{k \neq j} \frac{\lambda_k - \lambda_j - i c}{\lambda_k - \lambda_j + i c} = 1\, ,\quad j = 1,\, \cdots,\, N,
\ee
as we have discussed in the previous section.

From this analysis, we see the equivalence between the wave function of the 2D $\mathcal{N}=(2,2)^*$ $U(N)$ topological Yang-Mills-Higgs theory and the wave function of the (1+1)D quantum nonlinear Schr\"odinger equation in the $N$-particle sector. Hence, the duality between these two theories at quantum level is implied.

\subsection{Duality Web with Quantum KdV Equation}

Based on the discussions in Section~\ref{sec:duality} and \ref{sec:quKdV}, we can incorporate the quantum KdV equation into the duality discussed in Refs.~\cite{GS-1, GS-2}, which is briefly reviewed in the previous subsection. In the UV regime, the classical (m)KdV equation is dual to the classical nonlinear Schr\"odinger equation, while the quantum (m)KdV equation and the quantum nonlinear Schr\"odinger equation share the same Bethe Ansatz equation as the continuum limit of the spin-$\frac{1}{2}$ XXZ chain in the small anisotropy limit ($\eta \to 0$) or the continuum limit of the spin-$\frac{1}{2}$ XXX chain.

   \begin{figure}[!htb]
      \begin{center}
        \includegraphics[width=0.42\textwidth]{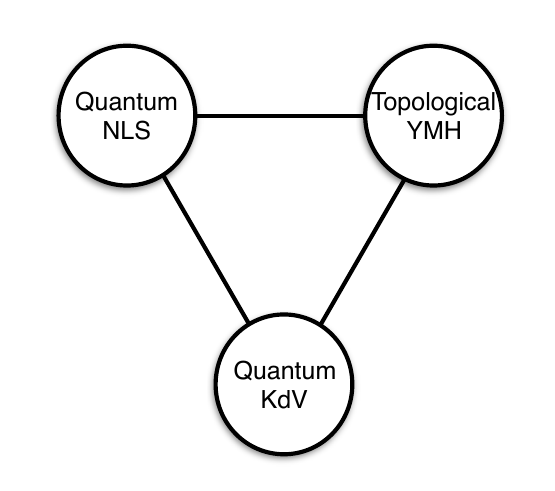}
      \caption{The duality web in the UV regime}
      \label{fig:rel}
      \end{center}
    \end{figure}

To summarize, in the UV regime we have not only the duality between the nonlinear Schr\"odinger equation and the 2D $\mathcal{N}=(2,2)^*$ topological Yang-Mills-Higgs theory, but also a duality web among the nonlinear Schr\"odinger equation, the KdV equation, the 2D $\mathcal{N}=(2,2)^*$ topological Yang-Mills-Higgs theory, with the duality between each of them. Schematically, the duality web can be shown in Fig.~\ref{fig:rel}.

\section{Discussion}\label{sec:discussion}

In this paper we discussed the relation between the (1+1)D nonlinear Schr\"odinger equation and the KdV equation both at the classical level and at the quantum level. We see that they share many properties especially in the UV regime. Some connections with the 2D $\mathcal{N}=(2,2)^*$ topological Yang-Mills-Higgs theory were also discussed.

There are many open problems that deserve more detailed research in the future. First, we would like to analyze the integrability of the full dual model obtained from the boson/vortex duality of the nonlinear Schr\"odinger equation. Also, besides the bulk theory, the matrix model obtained by including the boundary into the dual model is also worth studying. To understand this new matrix model can possibly deepen our previous knowledge on the KdV equation as a matrix model.

As we mentioned in Section~\ref{sec:quKdV}, it is still not very clear how to obtain the quantum KdV equation by taking an appropriate limit of the intermediate long wave equation (ILW). It would be very interesting to have a better understanding on this problem, which will consequently allow us to study the quantum KdV equation and understand its relation with the nonlinear Schr\"odinger equation in the gauge theory via the Bethe/Gauge correspondence.

It is known that a great amount of integrable models can be obtained from dimensional reduction of the 4D self-dual Yang-Mills theory \cite{Mason}. In particular, the nonlinear Schr\"odinger equation and the KdV equation belong to the same class in the dimensional reduction. This approach may provide us with a new perspective of the relation between these two theories, i.e., the correspondence can be understood in a geometrical way.

Moreover, it was suggested in Ref.~\cite{GS-1} to use the Nahm transform to understand the duality between the nonlinear Schr\"odinger equation and the topological Yang-Mills-Higgs theory. Based on the discussions in this paper, it is natural to expect that one can also repeat the Nahm transform analysis for the KdV equation, which we would like to investigate in the future.


\section*{Acknowledgements}

The author would like to thank Sasha Abanov, Ilmar Gahramanov, Partha Guha, Peter Koroteev, Yang Lan, Vasily Pestun, Nuno Rom\~ao, Vatche Sahakian, Pedro Vieira, Xinyu Zhang, Peng Zhao and Jian Zhou for many discussions. The author is also very grateful to Marco Rossi and Davide Fioravanti for comments.

\appendix

\bibliographystyle{utphys}
\bibliography{IntegModel}

\end{document}